\documentclass[entropy,article,submit,moreauthors,pdftex]{Definitions/mdpi}

\usepackage{quantikz}
\usepackage{amssymb}
\preto{\abstractkeywords}{\nolinenumbers}

\firstpage{1} 
\pubvolume{1}
\issuenum{1}
\articlenumber{0}
\pubyear{2022}
\copyrightyear{2022}
\history{Received: date; Accepted: date; Published: date}

\Title{Lossy Micromaser Battery: Almost Pure States in the Jaynes-Cummings Regime}

\newcommand{\be}{\begin{equation}}
\newcommand{\ee}{\end{equation}}

\Author{Vahid Shaghaghi$^{1,2,3}$\orcidA{}, Varinder Singh$^{3}$, Matteo Carrega$^{4}$\orcidE{}, Dario Rosa$^{3,5}$\orcidD{} and Giuliano Benenti$^{1,2,6}$\orcidC{}}

\AuthorNames{Vahid Shaghaghi, Varinder Singh, Giuliano Benenti, and Dario Rosa}

\address{%
$^{1}$ \quad  Center for Nonlinear and Complex Systems, Dipartimento di Scienza e Alta Tecnologia, Universit\`a degli Studi dell'Insubria, via Valleggio 11, 22100 Como, Italy \\
$^{2}$ \quad Istituto Nazionale di Fisica Nucleare, Sezione di Milano, via Celoria 16, 20133 Milano, Italy\\
$^{3}$ \quad Center for Theoretical Physics of Complex Systems, Institute for Basic Science (IBS), Daejeon - 34126, Korea \\
$^{4}$ \quad CNR-SPIN, Via Dodecaneso 33, 16146, Genova, Italy \\
$^{5}$ \quad Basic Science Program, Korea University of Science and Technology (UST), Daejeon - 34113, Korea \\
$^{6}$ \quad NEST, Istituto Nanoscienze-CNR, I-56126 Pisa, Italy
}

\corres{Correspondence: giuliano.benenti@uninsubria.it}

\abstract{We consider a micromaser model of a quantum battery, where the battery is a single mode of the electromagnetic field in a cavity, charged via repeated interactions with a stream of qubits, all prepared in the same 
non-equilibrium state, either incoherent or coherent, with the matter-field 
interaction modeled by the Jaynes-Cummings model. We show that the coherent
protocol is superior to the incoherent one, in that an effective pure
steady state is achieved for generic values of the model parameters.
Finally, we supplement the above collision model with cavity losses,
described by a Lindblad master equation. 
We show that battery performances, in terms of stored energy,
charging power, and steady-state purity, are slightly degraded up 
to moderated dissipation rate. Our results show that micromasers 
are robust and reliable quantum batteries, thus making them  
a promising model for experimental implementations.}


\begin{document}
\section{Introduction}
\label{sec:intro}


The description of the dynamics of open quantum systems via 
system-reservoirs collision models~\cite{Rau63, alicki_book,Scarani02,Ziman02,Ziman05,Benenti07,Giovannetti12,Uzdin14,Lorenzo15,Strasberg17,DeChiara18,Pezzutto19,Landi21,Shaghaghi22_PRE,perarnau_collisional,salvia2022quantum} is rapidly spreading 
in a variety of research areas, see~\cite{Campbell21, Ciccarello22} for reviews.
Notably in the growing field of quantum thermodynamics \cite{esposito09nonequilibrium, campisi11colloquium, anders15quantum, pekola21colloquium, carrega2022engineering}, it has been applied to address fundamental problems like the relaxation to equilibrium~\cite{Scarani02,Ziman02,Ziman05,Benenti07}, 
the relationship between information and thermodynamics~\cite{Lorenzo15}, the efficiency of 
thermodynamic heat engines~\cite{Strasberg17,DeChiara18,Pezzutto19}, exposed to either thermal or nonequilibrium 
reservoirs~\cite{Uzdin14,Pezzutto19,Shaghaghi22_PRE}, the relevance of non-Markovian effects~\cite{Giovannetti12,Pezzutto19} and 
strong system-bath coupling~\cite{Strasberg17}.

Recently, collision models have been used to describe the charging process 
of quantum batteries~\cite{Landi21,Shaghaghi22_PRE,perarnau_collisional,salvia2022quantum,Shaghaghi2022}, that is, of quantum mechanical systems suitable 
to store energy in some excited states~\cite{alicki_fannes_original, Hovhannisyan2013, Andolina2018,Zhang2019,Caravelli2020,Quach2020,Crescente2020, Friis2018,Rossini2019,Santos2019,Rosa2020, Binder2015,Campaioli2017,Le2018, Ferraro2018, Andolina2019, Crescente2020b, Ghosh2020, Rossini2020, GarciaPintos2020, hamma_comment, hamma_reply_to_comment,
Zakavati2021, Ghosh2021, Seah2021, mondal2022periodically, kanti2022quantum, gyhm2022quantum, salvia2022quantum, gemme2022IBM, mazzoncini2022optimal, erdman2022reinforcement}, 
to be later released on demand.
The charging of the battery, modeled by a quantum harmonic oscillator or a large spin, 
via sequential interactions (collisions) with a stream of qubits,
has shown enhanced performances with respect to classical counterpart, when the qubits are prepared in some 
coherent state, with respect to the classical, incoherent charging~\cite{perarnau_collisional,salvia2022quantum}. 

Moreover, it has been possible to show that by this charging protocol, and when the incoming qubits have a certain degree of coherence, the cavity reaches an effectively steady state which is both highly excited and essentially \textit{pure}.
Interestingly, these features extends beyond the weak coupling regime, and are present both in the strong and ultrastrong coupling regimes of field-matter interaction~\cite{Shaghaghi2022}.
Achieving a pure state is appealing in terms of ergotropy~\cite{Allahverdyan2004, Delmonte21, safranek2022work}, that is, 
of the amount of energy that can be extracted from a battery via 
unitary operations. While for a mixed state part of the excitation energy 
(measured from the ground state energy of the quantum battery) cannot be used,
for a pure state it is always possible to reach the ground state by a suitable unitary operation. 
Therefore for a pure state 
the ergotropy equals the mean excitation energy.
Moreover, the fact that an effective steady state is achieved 
forbids the cavity to absorb an unbounded amount of 
energy, a possibility that would be dangerous, 
conceptually equivalent to the burning of an overcharged 
classical battery.

In this paper, we show first of all that the battery stability is 
obtained only in the case of coherent charging, while in the incoherent 
case slight deviations from the ideal, fine-tuned conditions, can 
effectively burn the battery.
Moreover, the reported excellent properties, in the coherent case,  
of a micromaser as a quantum battery
call for a deep analysis of its stability in presence of decoherence mechanisms such as quantum noise~\cite{Ciccarello22, Carrega20, schnirman_2002, Devoret13, Calzona_2023}. 
Here, we consider the Jaynes-Cummings regime~\cite{meystre_book}, where counter-rotating terms in the
qubit-cavity collisions are neglected. Such approximation is naturally valid 
in the weak coupling regime, in which dissipative effects are small
during the time of a collision and the qubit-cavity coupling strength $g$ is 
much smaller than the cavity frequency $\omega$. Moreover, it can be pushed 
to the strong or ultra-strong coupling regime, $0.1\lesssim g/\omega\lesssim 1$, 
provided a simultaneous frequency modulation for both the qubit and the 
field is implemented~\cite{ultrastrong_JC}. 
We will model battery dissipation via a standard Lindblad master 
equation approach~\cite{Lindblad1976,GKS76,qcbook},
with losses due to the finite cavity lifetime.
While dissipation naturally avoids overcharging problems, forbidding 
unbounded energy absorption, its impact on the effectiveness of the
charging process, and in particular on the steady-state purity, is
a question to be carefully considered. 
In this paper, we show that the steady-state 
energy and purity are rather weakly affected up to $\gamma t_r=0.1$, with 
$\gamma$ being the cavity decay rate and $t_r$ the time interval between two consecutive 
collisions.

The paper is organized as follows. In Sec.~\ref{sec:model} we describe
the collision model here investigated. Incoherent and coherent chanrging
protocols are discussed in Sec.~\ref{sec:incoherent} and~\ref{sec:coherent}, 
respectively. In Sec.~\ref{sec:dissipation}, we consider the 
effects of dissipation, focusing on the coherent case, which is 
advantageous in the ideal, noiseless case.
Finally, our conclusions are drawn in Sec~\ref{sec:conclusions}.

\section{The model}
\label{sec:model}

The quantum battery (QB) that we consider consists of a quantized electromagnetic field in a cavity, which for our purposes can be modeled as a quantum harmonic oscillator.
The QB is initially prepared in the ground state, $\ket{0}$, of the harmonic oscillator.
Energy is then accumulated by sending into the cavity sequentially a stream of qubits (or two state systems), interacting with the harmonic oscillator and transferring energy into it.
The initial state of each qubit reads
\begin{align}
    \label{eq:initial_state_qubit}
    \rho_q &= q \ket{g} \bra{g} + (1 - q) \ket{e} \bra{e} 
+c \sqrt{q (1 - q)} \left( \ket{e} \bra{g} + \ket{g} \bra{e}\right) \, ,
\end{align}
where $\ket{g}$ is the ground state of the qubit while $\ket{e}$ represents the excited state. The parameters $q$ and $c$ control level populations and the degree of coherence, respectively.
The interaction between a qubit and the cavity field is described by the Hamiltonian \cite{meystre_book}
\begin{align}
    \label{eq:Rabi_hamiltonian}
    \hat{H} &= \hat{H}_0 + \hat{H}_1 \, , 
    \nonumber \\
    \hat{H}_0 & \equiv  \omega_F \hat{a}^\dagger \hat a + \frac 12 \omega_q \hat{\sigma}_z \, \nonumber \\
    \hat{H}_1 & \equiv g \left( \hat{a} \hat \sigma_+ + \hat{a}^\dagger \hat \sigma_- + \hat{a}^\dagger \hat \sigma_+ + \hat{a} \hat \sigma_- \right) \, ,
\end{align}
where $\omega_q$ and $\omega_F$ are the frequencies of the qubit and the field, respectively; $\hat{a}^\dagger$ and $\hat{a}$ are the creation/annihilation operators of the harmonic oscillator and $\hat{\sigma}_z$, $\hat{\sigma}_+$, $\hat{\sigma}_-$ denote the number, creation and annihilation operators for the qubit. 
Finally, $g$ is the coupling constant for the interaction.
We work in units such that $\hbar = 1$.

To describe the dynamics, it is convenient to move to the interaction picture. 
By further assuming that the qubits and the field are in \textit{resonance}, $\omega_q = \omega_F \equiv \omega$, the Hamiltonian simplifies to
\begin{equation}
    \label{eq:Rabi_hamiltonian_interaction}
    \hat{H}_I = g \left( \hat{a} \hat \sigma_+ + \hat{a}^\dagger \hat \sigma_- + e^{i 2 \omega t}\hat{a}^\dagger \hat \sigma_+ + e^{-i 2 \omega t} \hat{a} \hat \sigma_- \right) \, .
\end{equation}
The time evolution of the qubit-battery system is governed by the time-evolution operator
\begin{equation}
    \label{eq:time_evolution_op_def}
    \hat U_I(\tau) \equiv \mathcal{T} \exp{\left( - i \int_0^\tau \hat{H}_I(t) \mathrm d t \right)} \, ,
\end{equation}
where $\mathcal{T}$ is the time-ordering operator, 
and $\tau$ denotes the interaction time between a single incoming qubit and the harmonic oscillator.
The QB state after each collision can then be obtained by tracing out the qubit degrees of freedom.

 As it is well-known, under the weak-coupling regime --- $\frac g\omega \ll 1$ --- Eq.~\eqref{eq:time_evolution_op_def} effectively reduces to the Jaynes-Cummings evolution operator,
\begin{equation}
    \label{eq:jaynes_cummings_operator}
    \hat{U}_I = \exp{\left(- i \theta \left( \hat{a} \hat \sigma_+ + \hat{a}^\dagger \hat \sigma_- \right)\right)},
\end{equation}
where we have introduced the parameter $\theta \equiv g \tau$.

In the context of quantum batteries~\cite{Binder2015, Campaioli2017, gyhm2022quantum}, we are interested in fast charging and discharging, which requires going beyond the regime $\frac{g}{\omega} \ll 1$. While Eq.~\eqref{eq:jaynes_cummings_operator} is \emph{sic and simpliciter} 
no longer valid beyond this regime, we can, nevertheless, extend its validity into the strong coupling regime $0.1\lesssim \frac{g}{\omega}\lesssim 1$.
For that purpose, we take advantage of a simultaneous external frequency modulation of the qubit and field frequencies, as described in \cite{ultrastrong_JC}.
By modulating the frequencies we modify the Hamiltonian of the model to:
\begin{align}
    \label{eq:Rabi_hamiltonian_modulated}
    &\hat{H} \to \hat{H}^\prime  = \hat{H}_0^\prime + \hat{H}_1^\prime \, , \nonumber \\
    &\hat{H}_0^\prime  \equiv  \left(\omega_F + \eta \nu \cos(\nu t)\right) \hat{a}^\dagger \hat a + \frac 12 \left(\omega_q + \eta \nu \cos(\nu t)\right) \hat{\sigma}_z \, ,\nonumber \\
    &\hat{H}_1^\prime \equiv \hat H_1 \, ,
\end{align}
where $\eta$ and $\nu$ are the modulation amplitude and frequency, respectively. By a frequency modulation, the time evolution operator gets modified from Eq.~\eqref{eq:time_evolution_op_def} to Eq.~\eqref{eq:jaynes_cummings_operator}, which is therefore valid also in the ultrastrong coupling regime.

In the Jaynes-Cummings framework, the evolution of the battery state, $\rho_B$, can be described by the following master equation relating the state of the battery after $k + 1$ collisions to the state of the battery after $k$ collisions:
\begin{align}
    \label{eq:master_eq_general}
    \rho_B(k + 1) &= (1 - q) \hat{c}_{N + 1} \rho_{B}(k) \hat{c}_{N + 1} + q \hat{s}_{N + 1} \hat a \rho_{B}(k) \hat{a}^\dagger \hat{s}_{N + 1} + (1 - q) \hat{a}^\dagger\hat{s}_{N + 1} \rho_{B}(k) \hat{s}_{N + 1}\hat{a} + q  \hat{c}_{N} \rho_{B}(k) \hat{c}_{N} \nonumber \\
    & + i c \sqrt{q (1 - q)} \left[ \hat{c}_{N} \rho_{B}(k) \hat{s}_{N + 1} \hat a + \hat{c}_{N + 1} \rho_{B}(k) \hat{a}^\dagger \hat{s}_{N + 1} - \hat{s}_{N + 1} \hat a\rho_{B}(k)  \hat{c}_{N + 1} - \hat a^\dagger \hat{s}_{N + 1} \rho_{B}(k) \hat{a}^\dagger \hat{c}_{N} \right] \, ,
\end{align}
where the operators $\hat{c}_{N + 1}$, $\hat{c}_{N}$ and $\hat{s}_{N + 1}$ stand for $\cos\left(\theta \sqrt{\hat N + \hat 1} \right)$, $\cos\left(\theta \sqrt{\hat N } \right)$ and $\frac{\sin\left(\theta \sqrt{\hat N + \hat 1} \right)}{\sqrt{\hat N + \hat 1}}$, respectively.

The incoherent protocol corresponds to $c=0$, where the qubits are in a fully mixed state and without any degree of coherence. 
The other limiting case, $c=1$ \footnote{The effect of a complex phase in the coherence parameter, \textit{i.e.} $c \in \mathbb{C}$, has been considered in \cite{Shaghaghi2022}.}, corresponds to the fully 
coherent protocol, where the battery is charged by a stream of qubits 
prepared in a pure state, superposition of the states $|g\rangle$ and 
$|e\rangle$.

\section{Incoherent charging protocol}
\label{sec:incoherent}

Let us first focus on the case in which the incoming qubit is given by an incoherent mixture, \textit{i.e.} in which $c = 0$.
In this case, Eq.~\eqref{eq:master_eq_general} simplifies to 
\begin{align}
    \label{eq:master_eq_incoherent}
    \rho_{n}(k + 1) &= |s_{n + 1}|^2 \left( q \rho_{n+1}(k)  - (1-q) \rho_{n}(k)\right) 
    +  |s_{n}|^2 \left( (1 -q) \rho_{n-1}(k)  - q \rho_{n}(k)\right) + \rho_n(k)\, ,
\end{align}
where $\rho_n (k)$ denotes the diagonal elements of the system density matrix, $\rho_{nn}(k)$, and $s_{n} \equiv \sin(\theta \sqrt{n})$.

The main feature to observe in Eq.~\eqref{eq:master_eq_incoherent} is that the dynamics turns out to be peculiar when the parameter $\theta$ takes appropriate (fine-tuned) values of the form $\theta = \frac{ \pi}{\sqrt{m}}$, with $m$ being a positive integer.
To see this, let us consider first the case in which all the qubits are in the excited state, \textit{i.e.} $q = 0$, and  study the equation defining the steady state, $\rho_n (k + 1) = \rho_n (k)$ for each $n$.
It reads
\begin{equation}
    \label{eq:steady_state_eq_incoherent}
    |s_{n+ 1}|^2 \rho_n(k) - |s_{n}|^2 \rho_{n-1}(k) = 0 \, \qquad \forall n .
\end{equation}
When $\theta$ is fine-tuned to satisfy $\theta = \pi/\sqrt{m}$, the steady state, $\rho_n^{\mathrm{s.s.}}$, turns out simply to be 
\begin{equation}
    \label{eq:steady_state_micromaser_incoherent_fine_tuned}
    \rho_n^{\mathrm{s.s.}} = \delta_{n, m - 1} \, ,
\end{equation}
which means that the steady state is a \textit{pure state} and, more precisely, it is a number state.
We will refer to this property as \textit{trapping}.
In such a situation, the amount of extractable work from the battery turns out to be deterministic and maximal, since the ergotropy coincides with the total energy for pure states~\cite{Allahverdyan2004}.
The possibility of building number states out of a micromaser has been already discussed long ago in the literature \cite{Filipowicz_number_states}.
Here we use such an opportunity to build an incoherent quantum battery having the wanted properties of reliable and maximal work extraction.

To the purpose of building an effective device out of this property, one needs to be sure that such steady state can be reached in a finite (ideally small) number of collisions (thus increasing the charging power).
Moreover, it is crucial that such a trapping property is robust, meaning that it is not lost for deviations of the parameter $\theta$ from the fine-tuned value $\theta  = \pi/\sqrt{m}$, and for $q \ne 0$.

We start by considering the fine-tuned case at $q \neq 0$: It can be shown that the steady state equations are solved as follows:
\begin{align}
    \label{eq:steady_state_sol_incoherent_q_neq_0}
    & \rho_n^{\mathrm{s.s.}} = r^{n} \rho_{0}^{\mathrm{s.s.}} \,, \ \qquad \forall n  < m \, , \nonumber \\
    & \rho_n^{\mathrm{s.s.}} = 0\,, \qquad \forall n  \geq m \, , \nonumber \\
    & \rho_{0}^{\mathrm{s.s.}} = \frac{r - 1}{r^{m + 1} - 1} \, ,
\end{align}
where we defined $r \equiv \frac{1 - q}{q}$.
From Eq.~\eqref{eq:steady_state_sol_incoherent_q_neq_0} we see that the trapping property is essentially preserved for $q \neq 0$ and $\theta$ fine-tuned, since the energy of the steady state remains well-concentrated around the wanted value, $E_{m - 1}$, obtained for $q = 0$.
Given the steady state Eq.~\eqref{eq:steady_state_sol_incoherent_q_neq_0}, we can compute the purity, $\mathcal{P}$, of such state, defined as $\mathcal{P} \equiv \mathrm{Tr}\left( \rho^2 \right) = \sum_{n} \rho_{n}^2$, where the last equality uses the fact that in the incoherent case $\rho$ is diagonal.
We get
\begin{equation}
    \label{eq:purity_incoherent_sol_q_neq_0}
    \mathcal{P}^{\mathrm{s.s.}} = \frac{r - 1}{r + 1} \frac{(r^2)^{m + 1} - 1}{(r^{m + 1} - 1)^2} \, ,
\end{equation}
which, for $m$ sufficiently large, can be approximated by $\mathcal{P}^{\mathrm{s.s.}} \approx 1 - 2 q$.
Hence, we conclude that the purity property, enjoyed by the steady state for $q = 0$, is not stable and it is lost for $q \neq 0$. 

\begin{figure}[t!]
\begin{center}
    \includegraphics[width=1\hsize]{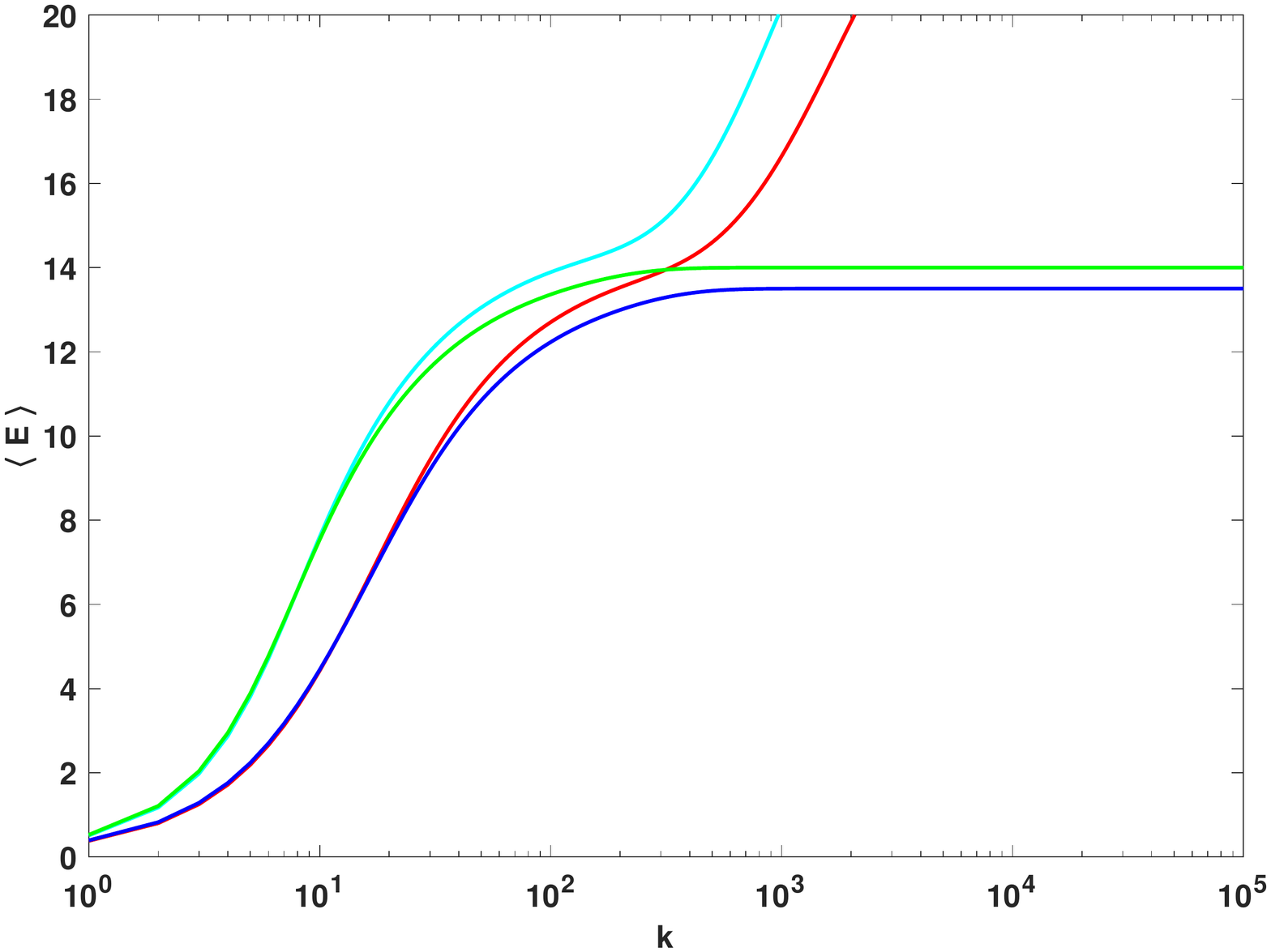}\\
    \caption{Energy of the battery vs. the number of battery-qubit collisions,
    for $q=0.25$ and $\theta=\frac{\pi}{\sqrt{15}}$ (blue line), $q=0.25$ and $\theta=\frac{\pi}{\sqrt{15.6}}$ (red line), $q=0$ and $\theta=\frac{\pi}{\sqrt{15}}$ (green line:), $q=0$ and $\theta=\frac{\pi}{\sqrt{15.6}}$ (cyan line).}
\label{fig:incoherent_energies}
\end{center}
\end{figure}

On the other hand, we see that the trapping property itself is not stable for deviations of $\theta$ from its fine tuned value. 
In such a case, the steady state equations do not admit any trapping solution and the battery levels are populated without any upper bound.
In Fig.~\ref{fig:incoherent_energies} we show the energy stored in the battery, as a function of the number of collisions for both fine-tuned values of the $\theta$ and $q$ parameter and for small deviations around their fine-tuned values.
We find, in agreement with the previous analytical discussion, that the battery, when $\theta$ and $q$ are fine-tuned, reaches the steady state $\ket{m}$.
On the other hand, we also observe that the system is not robust to small perturbations of $\theta$.
In such cases, a temporary, \textit{metastable}, state is reached at first, 
for longer times as $\theta$ approaches a fine-tuned value, 
followed by a further evolution which continues indefinitely.
In addition, for fine-tuned values of $\theta$ but $q \neq 0$, the trapping properties are robust and the energy of the steady state is close to its fine-tuned counterpart.

\begin{figure}[t!]
\begin{center}
    \includegraphics[width=1\hsize]{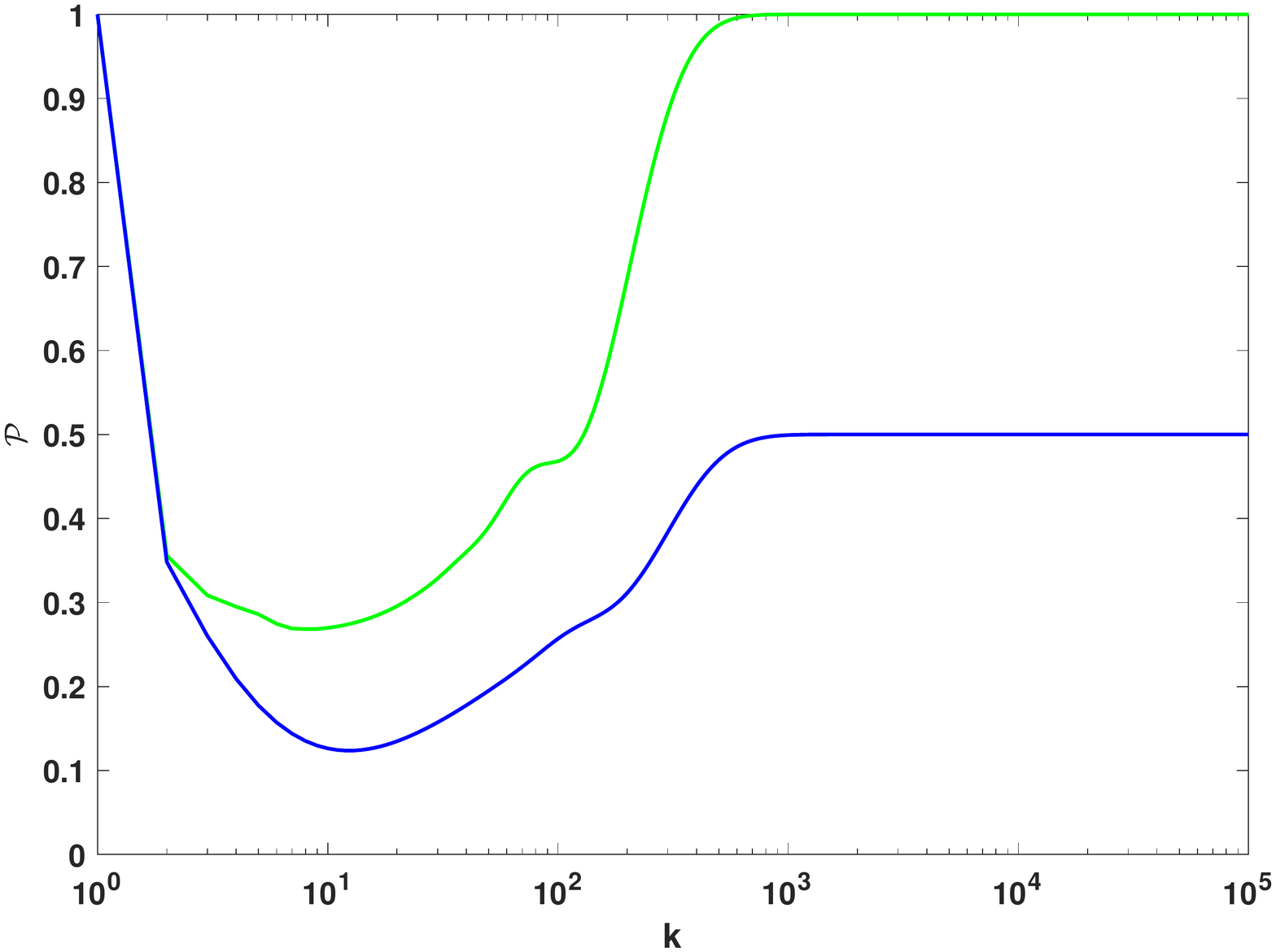}\\
    \caption{Purity of the battery vs. the number of battery-qubit collisions,
    for the fine-tuned value $\theta=\frac{\pi}{\sqrt{15}}$, $q=0.25$ 
    (blue line) and $q=0$ (green line).}
\label{fig:incoherent_purities}
\end{center}
\end{figure}

Since fine-tuned values of the $\theta$ parameter lead to stable energies, we have studied in these cases the purity of the battery state $\rho(k)$, as a function of the number of collisions. The results are reported in Fig.~\ref{fig:incoherent_purities}.
As expected, in the case of fine-tuned $\theta$ values, the final state has purity $1$ when $q = 0$, but such a property is lost for $q \neq 0$.  

Summarising, we have shown that in the case of incoherent charging the fine tuned steady states $\ket{m}$ are fragile against small perturbations of both 
the $\theta$ and $q$ parameters.

\section{Coherent charging protocol}
\label{sec:coherent}

For completeness and ease of comparison with the incoherent case, let us now
consider the case in which the qubits are coherent, \textit{i.e.} $c \neq 0$ in Eq.~\eqref{eq:master_eq_general}, which was the subject of Ref.~\cite{Shaghaghi2022}.
We will focus on the case $c = 1$, which means that the incoming qubits are in a superposition state: 
\begin{align}
\label{eq:in_state_qubit_coherent}
  \rho_q = \ket{\psi} \bra{\psi} \, , \quad  \ket{\psi} \equiv \sqrt{q} \ket{g} + \sqrt{1 - q} \ket{e}  \, . 
\end{align}

In this case, it has already been discussed in the literature \cite{slosser_PRL, slosser_PRA} that the purity properties of the trapping states discussed in Sec.~\ref{sec:incoherent} are preserved for perturbations of the $q$ parameter.
In other words, it has been shown analytically that the micromaser reaches a pure steady state whenever the $\theta$ value only is fine tuned, \textit{irrespective} on the value of $q$:
\begin{equation}
    \label{eq:purity_cond_coherent_slosser}
    \theta = \frac{Q}{\sqrt{m}}\,\pi , \qquad \forall \,Q,m \in \mathbb{N} \,\; \mathrm{and} \,\; \forall q \in \left[0 , 1 \right] \, .
\end{equation}

\begin{figure}[t!]
\begin{center}
    \includegraphics[width=1\hsize]{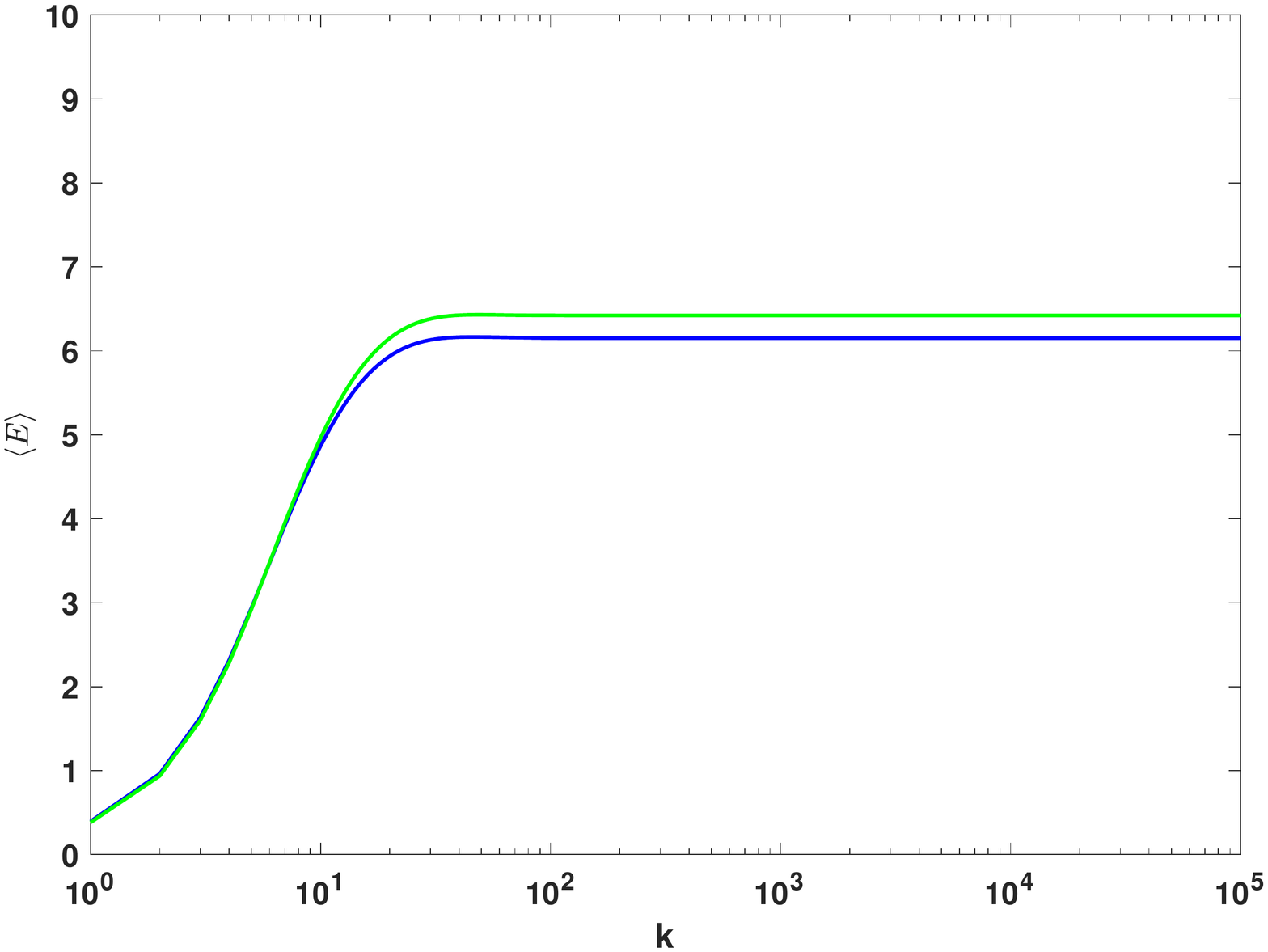}\\
    \caption{Energy of the battery vs. the number of battery-qubit collisions,
    for $q=0.25$, $\theta=\frac{\pi}{\sqrt{15}}$ (blue line) and $\theta=\frac{\pi}{\sqrt{15.6}}$ (green line).}
\label{fig:coherent_energies}
\end{center}
\end{figure}

\begin{figure}[t!]
\begin{center}
    \includegraphics[width=1\hsize]{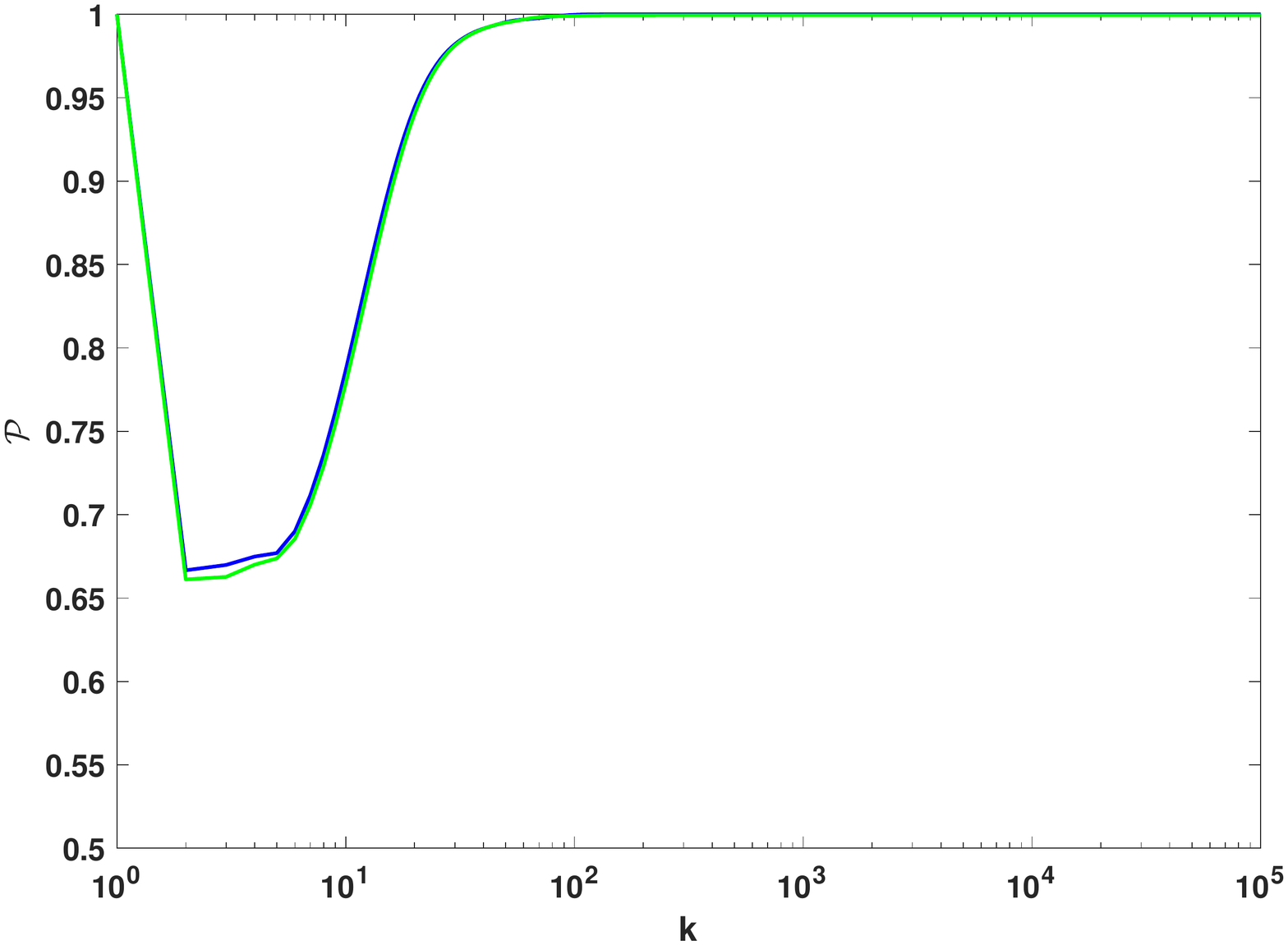}\\
    \caption{Purity of the battery vs. the number of battery-qubit collisions,
    for the same parameter values as in Fig.~\ref{fig:coherent_energies}.}
\label{fig:coherent_purities}
\end{center}
\end{figure}

On the other hand, the robustness of such steady states for perturbations of the $\theta$ value has been discussed only recently in Ref.~\cite{Shaghaghi2022}.
In view of the discussion done in the previous section, this is the most interesting case to consider, since the steady state itself is fragile for perturbations of the $\theta$ parameters in the incoherent case $c = 0$.
In Fig.~\ref{fig:coherent_energies} and \ref{fig:coherent_purities}, 
we show data for energy and purity during the charging protocol 
of the micromaser, when the incoming qubits are fully 
coherent, that is, $c = 1$.

Quite remarkably, and in sharp contrast 
with the incoherent case $c=0$, we observe that the system reaches long-lived states for all the values of $\theta$
and $q$, and such states have the wanted property of being pure up to our best numerical precision~\cite{Shaghaghi2022}.
In other words, we see that thanks to the coherent charging protocol, our quantum battery reaches an almost steady state which is self-sustained by its own dynamics and from which all the energy stored can in principle be extracted, since it is effectively a pure state.
While such state may be a metastable one~\cite{Shaghaghi2022}, still 
our extensive simulations show their stability up to the time 
scales accessible to numerical investigations.

Another important feature to discuss of this charging protocol is that the time scale to saturate the energy in the battery coincides with the time scale to reach the desired unit purity. 
In the incoherent (fine-tuned) case, instead, these two time scales are 
separated by essentially an order of magnitude
(see Fig.~\ref{fig:incoherent_energies}, where the energy saturates after about 300 collisions, and Fig.~\ref{fig:incoherent_purities}, where the unit purity 
requires almost 1000 collisions to be achieved).

On the other hand, we observe that the final energy is significantly lower when compared to the incoherent setup.
The reason for this reduction in the stored energy can be studied analytically for the case of $\theta$ fine tuned ($\theta = \frac{\pi}{\sqrt{m}}$), discussed in \cite{slosser_PRL}.
In this case, the populations $\rho_{n}$ of the steady state satisfy the following set of recursion relations:
\begin{equation}
    \label{eq:recursion_coherent_fine_tuned}
    \rho_n = \frac{1 - q}{q} \mathrm{cot}^2\left(\frac{\pi}{2\sqrt{m}} \sqrt{n}\right) \rho_{n-1}\,, \quad \forall n<m \, ,
\end{equation}
from which one can observe that contrary to the incoherent case, the additional $n$-dependent terms $\mathrm{cot}^2\left(\frac{\pi}{2\sqrt{m}} \sqrt{n}\right)$ suppress the high energy populations and consequently reduces the energy stored in the battery.

\section{Lossy cavity}
\label{sec:dissipation}

In this section we analyze how much the features uncovered so far for the coherent charging protocol ($c = 1$) are stable when considering dissipative effects in the micromaser.
To this purpose, we consider the following master equation, where Eq.~\eqref{eq:master_eq_general} is modified by adding a dissipative term \cite{meystre_book}
\begin{align}
    \label{eq:master_eq_dissipative}
    \rho_B(k+1)=e^{\mathcal{L} t_r}{\rm Tr}_q[\hat{U}_I(\tau) (\rho_B(k)\otimes\rho_q)\hat{U}_I^\dagger(\tau)],
\end{align}
where $t_r$ denotes the time interval between consecutive collisions, 
assumed to be much larger than the duration $\tau$ of a single collision, so that 
at most a single qubit at a time interacts with the cavity.
We also assume that dissipation takes place only between two collisions.
Such an assumption is reasonable whenever the condition, \(t_r \gg \tau\) is satisfied.
For the sake of simplicity we have assumed the constant dissipation rate 
$\gamma$ to be sufficiently small ($\gamma \tau\ll 1$) 
to allow a single-step Trotterization of collision and dissipation,
this latter ruled by the Lindbladian $\mathcal{L}$. 
Dissipation of the cavity field is described by the master equation for a damped harmonic oscillator, with the Lindbladian~\cite{Gardiner_book}
\begin{align}
    \label{eq:damped harmonic oscillator}
 \mathcal{L}(\rho_B)=-\frac{\gamma}{2}(\Bar{n}+1)[a^\dagger a\rho_B-a\rho_B a^\dagger]-\frac{\gamma}{2}\Bar{n}[\rho_B a a^\dagger-a^\dagger\rho_B a]+H.c.\,.
\end{align}
In the equation above, $\Bar{n}$ denotes the mean number of thermal photons inside the cavity at a given temperature, and $\gamma$ stands for the decay rate of the mean number of photons in the cavity field towards $\Bar{n}$.

Note that in the fine-tuned case, $\theta = \frac{\pi}{\sqrt{m}}$, it was observed numerically in \cite{slosser_dissipation} that, although the separation of the phase space in disconnected sectors is no longer at play in the dissipative case, still the model exhibits long-lived states resembling their non-dissipative counterparts.
In particular, it was observed that such states are almost, but not completely, pure.

\begin{figure}[t!]
\begin{center}
    \includegraphics[width=1\hsize]{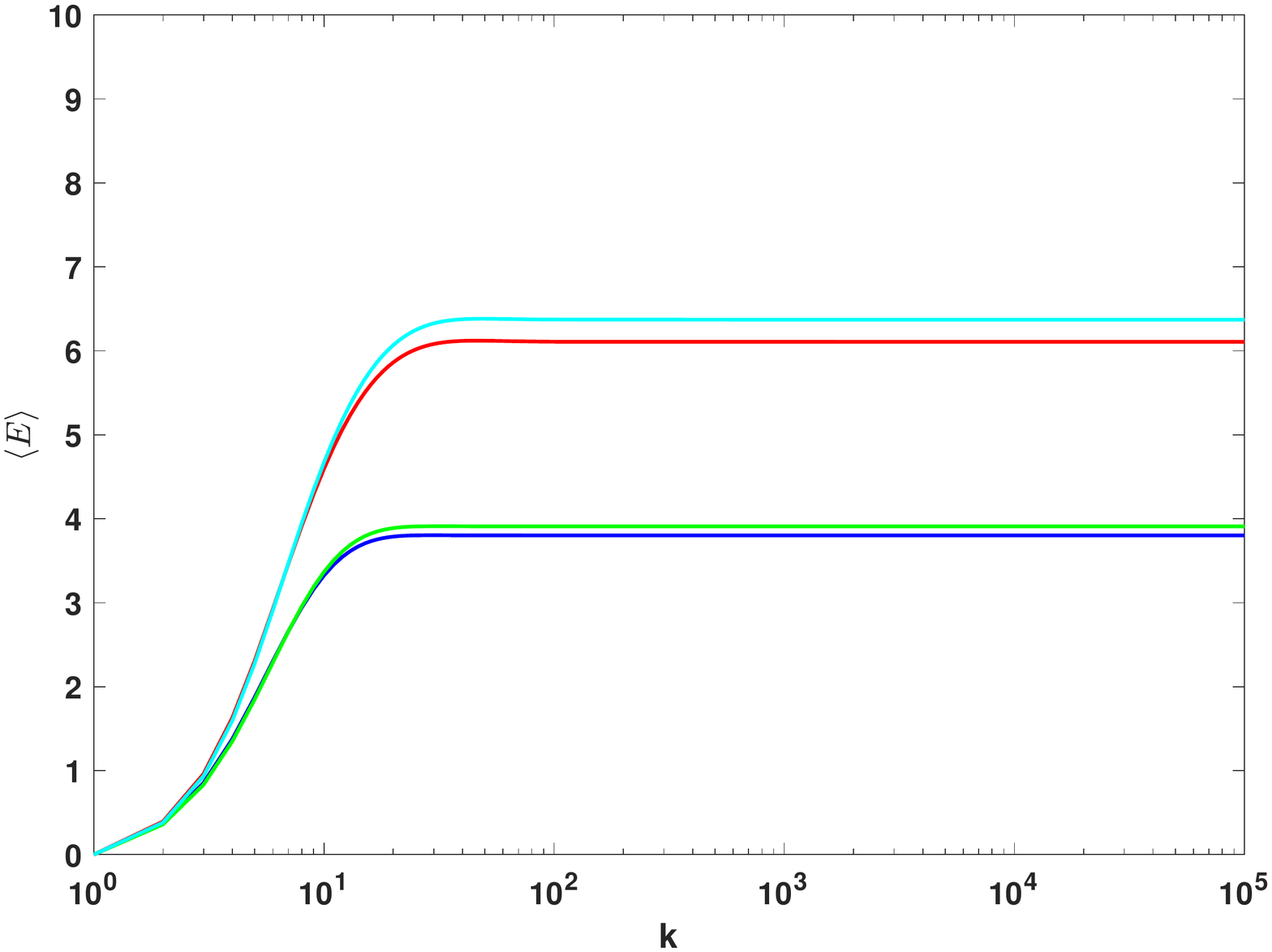}\\
    \caption{Energy of the battery vs. the number of battery-qubit collisions, in the presence of dissipation, at $q=0.25$ and  $\Bar{n}=0.15$,
    $\theta=\frac{\pi}{\sqrt{15}}$ and $\gamma t_r=0.1$ (blue line),
    $\theta=\frac{\pi}{\sqrt{15.6}}$ and $\gamma t_r=0.1$ (green line),
    $\theta=\frac{\pi}{\sqrt{15}}$ and $\gamma t_r=0.001$ (red line),
    $\theta=\frac{\pi}{\sqrt{15.6}}$ and $\gamma t_r=0.001$ (cyan line).}
\label{fig:dissipative_energies}
\end{center}
\end{figure}

\begin{figure}[t!]
\begin{center}
    \includegraphics[width=1\hsize]{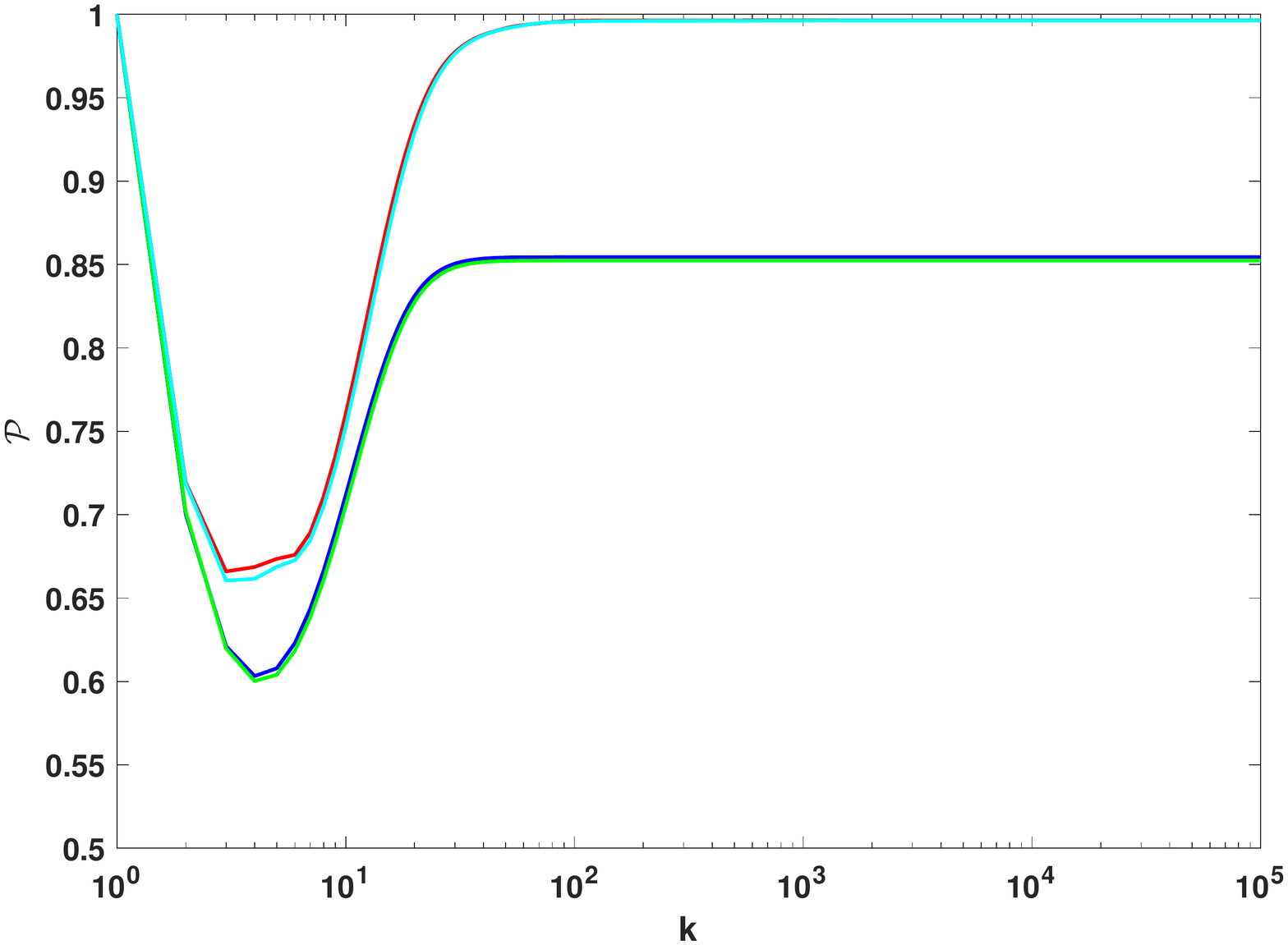}\\
    \caption{Purity of the battery vs. the number of battery-qubit collisions, 
    for the same parameter values as in Fig.~\ref{fig:dissipative_energies}.}
\label{fig:dissipative_purities}
\end{center}
\end{figure}

We have solved numerically Eq.~\eqref{eq:master_eq_dissipative} for $\gamma t_r \ll 1$ and $\Bar{n}=0.15$. 
In Fig.~\ref{fig:dissipative_energies}, we show the time evolution of the energy stored in the battery in presence of dissipation for different values of $\theta$.
As we see, for a low decay rate, $\gamma t_r=10^{-3}$, the system reaches a stable value of the energy which is almost the same as the energy stored in the absence of dissipation. For a high decay rate, $\gamma t_r=0.1$, the system again reaches a stable value of the energy, but with lower energy and charging power
compared to the energy stored in the non-dissipative case.
As a matter of fact one can also note in Fig.~\ref{fig:dissipative_purities} that for the case $\gamma t_r=10^{-3}$ the final state is almost pure and stable under its own dynamics. Furthermore, for the case $\gamma t_r=0.1$, purity drops to a lower value, $\mathcal{P}\approx 0.85$, but its stability is preserved.

The results of this section show that, even in the presence of a certain degree of dissipation, the qualitative features of the charging protocol in presence of coherences are preserved, with the final state being essentially pure and stable under time evolution.
These findings reinforce the idea that micromasers can be used to practically realize quantum batteries, showing excellent performances in terms of the stability of the accumulated energy and of the amount of energy that can be in principle extracted under unitary operations.

\begin{figure}[t!]
\begin{center}
    \includegraphics[width=1\hsize]{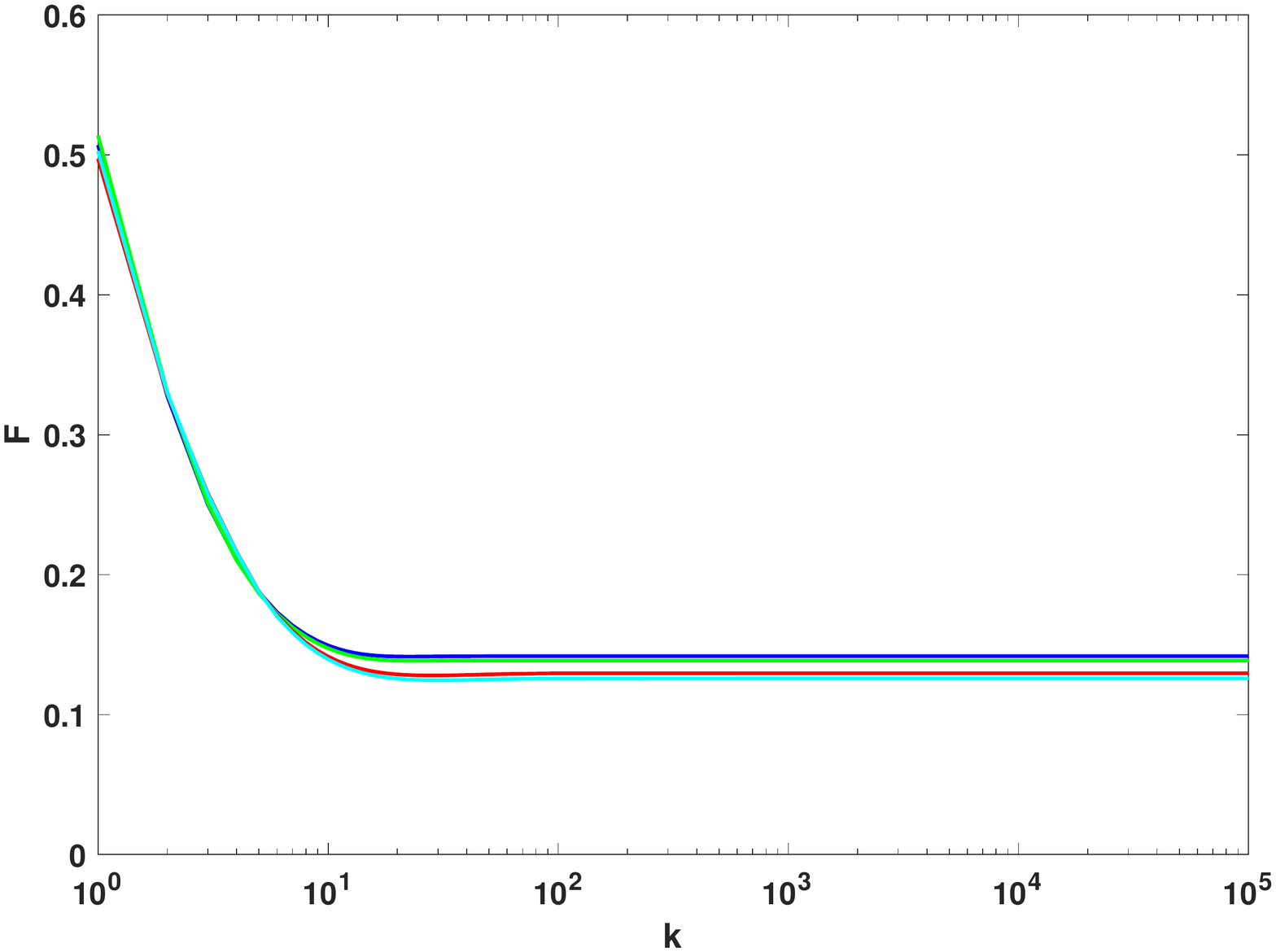}\\
    \caption{Fano factor vs. the number of battery-qubit collisions, in the presence of  dissipation, for $q=0.25$ and $\Bar{n}=0.15$, $\theta=\frac{\pi}{\sqrt{15}}$ and $\gamma t_r=0.1$ (blue line), $\theta=\frac{\pi}{\sqrt{15.6}}$ and $\gamma t_r=0.1$ (green line), $\theta=\frac{\pi}{\sqrt{15}}$ and $\gamma t_r=0.001$ (red line), $\theta=\frac{\pi}{\sqrt{15.6}}$ and $\gamma t_r=0.001$ (cyan line).}
\label{fig:fano_factor}
\end{center}
\end{figure}

Finally, we investigate the statistics of the field in the lossy cavity. In order to characterize the nature of photon statistics, we use Fano factor defined as
\begin{align}
    \label{eq:Fano_factor}
    F(k)=\frac{\sigma^2(k)}{\mu(k)},
\end{align}
where $\sigma^2(k)$ and $\mu(k)$ are the variance and the mean of the photon distribution after $k$ collisions. For $F>1$, $F=1$, and $F<1$ the cavity field has super-Poissonian, Poissonian, and sub-Pissonian statistics. Fig.~\ref{fig:fano_factor} shows the Fano factor as a function of the 
number of collisions, for the lossy cavity. We see that the state of the cavity field for different values of the decay rate follows the sub-Poissonian statistics, which is a property of non-classical fields. 

\section{Conclusions}
\label{sec:conclusions}

In this work we have further investigated the proposal of Ref.~\cite{Shaghaghi2022} that a micromaser, charged by means of coherent qubits, 
works as a very reliable quantum battery system. In contrast with incoherent charging, an effectively steady state, stable up to the times accessible to numerical investigations, is obtained also for generic values of $\theta=g\tau$. In contrast, in the incoherent case, for a generic and non fine-tuned value of $\theta\ne \pi/\sqrt{m}$ the battery absorbs an unlimited amount of energy, which would eventually ``burn it up''. We have also shown that 
our results are stable up to moderate values of battery damping, resulting in only a moderate reduction in energy and purity for the numerically observed steady state.

Our analysis has been limited to the Jaynes-Cummings regime, which naturally features in the original micromaser experiments~\cite{meystre_book}. On the other hand, one can address the strong or even ultra-strong coupling regime,
where the qubit-cavity interaction energy becomes comparable, or can even exceed the bare frequencies of the uncoupled systems~\cite{FornDiaz2019,Kockum2019}, and in such regime
counter-rotating terms in the interaction Hamiltonian 
should also be considered with care. 
Such terms can be handled using a simultaneous modulation 
of cavity and qubit frequency, as discussed in the literature~\cite{ultrastrong_JC}. 
However, such modulation requires a further control knob, 
and the stability of battery charging under unavoidable imperfections
in the modulation protocol should be investigated. 
Therefore, it would be interesting to investigate the working of 
the battery including counter-rotating terms, together with cavity losses
that might stabilize the battery even in this case. 
A careful examination of the above outlined strategies is worthwhile, 
since the ultra-strong coupling regime in principle offers the possibility of 
speeding up battery charging, and in general quantum operations.

In perspective it would be interesting to inspect solid state platforms for micromaser-based QBs, exploiting semiconducting double quantum dot geometry~\cite{liu2015semiconductor} or superconducting quantum circuits~\cite{you2007persistent}.

On a more fundamental ground, it would be interesting to consider a network 
of cavities and see whether quantum correlations between the batteries might
speed up the charging process. Moreover, it is crucial to design suitable 
discharge protocols to quickly and reliably extract the energy stored in the battery.

%
%

\authorcontributions{
Vahid Shaghaghi performed numerical simulations. 
The work was supervised by G.B., with inputs from D.R. M.C. and Varinder Singh.
All authors discussed the results and contributed to writing and revising the manuscript. 
}

\funding{Vahid Shaghaghi, Varinder Singh and D. R. acknowledge support from the Institute for Basic Science in Korea (IBSR024-D1). G.B. acknowledges financial support by the Julian Schwinger Foundation (Grant JSF-21-04-0001) and by INFN through the project QUANTUM.}

\conflictsofinterest{The authors declare no conflict of interest.
The funders had no role in the design of the study; in the collection, analyses, or interpretation of data; in the writing of the manuscript, or in the decision to publish the results.}



\externalbibliography{yes}
\bibliography{main.bib}






\end{document}